%% file: main.tex
\documentclass[runningheads,a4paper]{llncs}

\usepackage{amssymb}
\usepackage{graphicx}
\usepackage{subfigure}
\usepackage{color}
\usepackage{listings}

\usepackage{url}
\newcommand{\keywords}[1]{\par\addvspace\baselineskip
\noindent\keywordname\enspace\ignorespaces#1}

\input{ines}

\def\vtable{{\tt vtable}}
\def\virtual{{\tt virtual}}

\begin{document}

\mainmatter  

\title{Wiselib: A Generic Algorithm Library for Heterogeneous Sensor
  Networks}
\authorrunning{Baumgartner, Chatzigiannakis, Fekete, Koninis,
  Kr{\"o}ller, Pyrgelis}

\author{
Tobias Baumgartner\inst{1}
  \and Ioannis Chatzigiannakis\inst{2,3} 
  \and S{\'a}ndor Fekete\inst{1}
  \and \\Christos Koninis\inst{2,3} 
  \and Alexander Kr{\"o}ller\inst{1} 
  \and Apostolos Pyrgelis\inst{3}
}

\institute{
Braunschweig Institute of Technology, IBR, Algorithms Group,
    Germany\\
  \and Research Academic Computer Technology Institute, Patras, Greece\\
  \and Computer Engineering and Informatics Department, University of Patras,
    Greece\\
  Email:\email{ \{t.baumgartner, s.fekete, a.kroeller\}@tu-bs.de,\\
    \{ichatz, koninis\}@cti.gr, pyrgelis@ceid.upatras.gr}
}

\maketitle

\begin{abstract}
  One unfortunate consequence of the success story of wireless sensor
  networks (WSNs) in separate research communities is an ever-growing
  gap between theory and practice. Even though there is a increasing
  number of algorithmic methods for WSNs, the vast majority has never
  been tried in practice; conversely, many practical challenges are
  still awaiting efficient algorithmic solutions. The main cause for
  this discrepancy is the fact that programming sensor nodes still
  happens at a very technical level. We remedy the situation by
  introducing {\em Wiselib}, our algorithm library that allows for
  simple implementations of algorithms onto a large variety of
  hardware and software. This is achieved by employing advanced C++
  techniques such as templates and inline functions, allowing to write
  generic code that is resolved and bound at compile time, resulting
  in virtually no memory or computation overhead at run time.

  The Wiselib runs on different host operating systems, such as
  Contiki, iSense OS, and ScatterWeb. Furthermore, it runs on virtual
  nodes simulated by Shawn. For any algorithm, the Wiselib provides 
  data structures that suit the specific properties of the target
  platform. Algorithm code does not contain any platform-specific
  specializations, allowing a single implementation to run natively on
  heterogeneous networks.

  In this paper, we describe the building blocks of the Wiselib,
  and analyze the overhead. We demonstrate the effectiveness of
  our approach by showing how routing algorithms can be
  implemented. We also report on results from experiments with real
  sensor-node hardware.

\end{abstract}

\input{sec_intro}
\input{sec_rw}
\input{sec_problem}

\input{sec_wiselib}
\input{sec_case_study}
\input{sec_results}
\input{sec_access}
\input{sec_conclusion}

\subsubsection*{\ackname}

This work has been partially supported by the European Union under
contract number ICT-2008-224460 (WISEBED).

\bibliographystyle{abbrv}
\bibliography{main}

\end{document}

%% file: ines.tex
{\end{list}}

{\end{list}}

{\end{list}}

{\end{list}}

{\end{list}}

{\end{list}}

\newenvironment{dense_enumerate_1}{%
\begin{list}{\arabic{enumi}.}%
{\usecounter{enumi}%
\setlength{\topsep}{1mm}%
\setlength{\partopsep}{0mm}%
\setlength{\parskip}{0mm}%
\setlength{\parsep}{0mm}%
\setlength{\itemsep}{0mm}%
\settowidth{\labelwidth}{(1)}%
\setlength{\leftmargin}{0mm}%
\addtolength{\leftmargin}{\labelwidth}%
\addtolength{\leftmargin}{\labelsep}%
\setlength{\itemindent}{0mm}}}%
{\end{list}}

{\end{list}}

{\end{list}}

{\end{list}}

{\end{list}}

{\end{list}}


%% file: sec_intro.tex
\section{Introduction}
\label{sec:intro}

Since the initial visions proposed in the SmartDust
project~\cite{smartdust} ten years ago, Wireless Sensor Networks have
seen a tremendous development, both in theory and in practice.  On the
practical side, we see working sensor networks and applications in
many areas, from academia to industrial appliances. There is a large
variety of hardware and software to choose from that is easy to set up
and use.

This success story has also led to a serious practical issue that has
not been sufficiently addressed in the past: Sensor node brands are
very different in their capabilities. Some nodes have 8-bit
microprocessors and tiny amounts of RAM, while others burst with
power, being able to run desktop operating systems such as
Linux. Consequently, the software running on these systems is very
different on the various nodes. While it is easy to write code for a
specific platform, it is a very challenging task to develop
platform-independent code. Even worse, the operating systems on most
sensor nodes provide barely enough functionality to implement simple
algorithms. This means that the developer is forced to spend great
attention on low-level details, making the process painfully complex
and slow.

A parallel success story can be observed on the theoretical side, where
the development of distributed algorithms for many actual or hypothetical
problems has grown into a research field of its own. This has led to
a large variety of highly sophisticated algorithms for all kinds of tasks.
Unfortunately, many of them have never been tried in practice, due to
the overly difficult implementation process. Where algorithms are
implemented, they are hard to share and compare, as implementations
cannot be easily ported to new platforms.  Moreover, many important
challenges are not even addressed, as they can only be identified and
resolved by close collaboration between theory and practice.

This growing gap between theory and practice forms a major impediment
for exploiting the possibilities of complex distributed systems. The
{\em Wiselib} is our proposal to remedy this unfortunate situation. We
present a framework, written in C++, for platform-independent
algorithm development. Each algorithm written for the Wiselib can be
compiled for any supported system without changing any line of
code. It provides simple interfaces to the algorithm developer, with a
unified API and ready-to-use data structure implementations. The {\em
  Wiselib} addresses the following issues:

\smallskip\noindent\textbf{Platform independence.}  Wiselib code can
be compiled on a number of different hardware platforms, usually
without platform-dependent configurations, i.e., no ``{\tt \#ifdef}''
constructions. See Section~\ref{sec:problem:hetero} for details.

\smallskip\noindent\textbf{OS independence.} Wiselib code can be
compiled for different operating systems. This includes systems based
on C like Contiki, as well as C++ (the iSense firmware) and nesC
(TinyOS).

\smallskip\noindent\textbf{Exchangeability.} Algorithms and
applications can be composed of different components that interact
using well-defined interfaces, called {\em concepts}. Components can
be exchanged with other implementations without affecting the
remaining code. Moreover, both generic components and highly optimized
platform-specific components can be used simultaneously.

\smallskip\noindent\textbf{Broad algorithm coverage.} The Wiselib
currently covers a large variety of algorithms. It will contain
algorithms for each of the following categories:
  \noindent\parbox[t]{.5\textwidth}{
    \begin{dense_enumerate_1}
    \item routing algorithms
    \item clustering algorithms,
    \item time-synchronization algorithms,
    \end{dense_enumerate_1}
  }\hfill\parbox[t]{.45\textwidth}{
    \begin{dense_enumerate_1}\setcounter{enumi}{3}
    \item localization algorithms,
    \item data dissemination, and
    \item target tracking.
    \end{dense_enumerate_1}
  }

\smallskip\noindent\textbf{Cross-layer algorithms.}  In Wiselib an
algorithm can be designed to use other algorithm concepts, thus
enabling the use of existing algorithms for the implementation of more
complex ones. Moreover, we can stack protocols on top of each other,
extending their functionality.  See Section~\ref{sec:case_study} for
details.

\smallskip\noindent\textbf{Standard compliance.} The library is
written in a well-defined language subset of ISO~C++. This has a
number of benefits over custom languages such as nesC: The compilers
are more mature and better supported, and there is a large user base
that knows C++ from desktop development.

\smallskip\noindent\textbf{Scalability and efficiency.} The Wiselib is
capable of running on a great variety of hardware platforms, with CPUs
ranging from 8-bit microcontrollers to 32-bit RISC CPUs, and with
memory ranging from a few kilobytes to several megabytes. Algorithms
need to be very resource-friendly on the platforms from the lower end,
and at the same time be able to use more resources if available.

To our knowledge, the Wiselib is the only successful attempt to
achieve all of these goals at once. In this paper, we present the
basic building-blocks of the Wiselib, and show that the flexibility of
the design has barely any overhead---neither in code size nor in
run-time; one can simply add new algorithms only by following the
presented approach using the Wiselib interfaces. The algorithm can
then run on each supported sensor node or simulation platform. Our
goal is to achieve a state in which such an algorithm runs on
heterogeneous sensor networks, and even more, networks in which some
parts consist of virtual nodes running in a simulator.


This paper is organized as follows: The next section provides an
overview of related work, covering competing approaches as well as
implementations that inspired this work. Section~\ref{sec:problem}
explores the problem space by discussing the target platforms on which
we wish to run the Wiselib. Section~\ref{sec:wiselib} presents details
on the design of the Wiselib. In Section~\ref{sec:case_study} we
describe example implementations of routing algorithms; in
Section~\ref{sec:results}, we report on the surprisingly small code
and memory footprint on different platforms. Section~\ref{sec:access}
describes the current distribution of the Wiselib. We conclude the
paper in Section~\ref{sec:conclusion}.


%% file: sec_rw.tex
\section{Related Work}
\label{sec:related_work}

Efficient algorithm libraries have a long-standing tradition on
desktops and servers. The three libraries that motivated our work are
the Standard Template Library (STL), the Computational Geometry
Algorithms Library (CGAL)~\cite{cgal}, and Boost~\cite{boost}. They
share a great programming concept that we heavily use for the Wiselib:
Using C++ templates, one can construct complex object-oriented
software architectures that can be parameterized for many different
applications. The price of generality is paid at compile time. The
final binary contains highly efficient and specialized code, so that
there is no overhead at runtime.

The situation in sensor networks is not as promising. There have been
approaches to overcome the issues of incompatible nodes by providing
generic operating systems that run on multiple platforms. Examples are
Contiki~\cite{dunkels04contiki} and TinyOS~\cite{tinyos}. Neither runs
on all platforms we are envisioning for the Wiselib. Even worse, both
introduce new programming paradigms that are valid only for the
specific targets, such as protothreads in Contiki, and the whole
programming language nesC~\cite{nesc03} of TinyOS. The C-inspired nesC
attempts to allow for the construction of component architectures with
early binding, similar to the Wiselib, but achieves this through
introducing a new language that requires a custom compiler.

A challenging issue are heterogeneous networks. It is very simple to
have nodes exchange messages if they are of the same kind, and with
the same operating systems. It becomes surprisingly hard to let nodes
of different brands communicate with each other, even if both of them
use standardized IEEE 802.15.4 radios. A promising approach is the
Rime Stack~\cite{dunkels07-adaptive_communication,rimestack}, a
layered communication stack for sensor networks. It runs only on Contiki.
Recently, Sauter et al.~\cite{sauter09} demonstrated that is is
possible to communicate between sensor nodes running Contiki and
TinyOS. Since TinyOS uses IEEE 802.15.4, the Rime Stack and Chameleon
Module had been modified on Contiki.

Another attempt to produce a well-defined environment that runs on
different platforms was proposed by Boulis et
al.~\cite{boulis03-framework_scripting}: SensorWare defines a custom
scripting language; its syntax is based on Tcl. Consequently it
focuses on richer platforms with at least 1 Mbyte of ROM and 128
KBytes of RAM. A similar approach is Mat{\'e}~\cite{levis02-mate}, a
virtual machine running on top of TinyOS. It targets also small
devices with a very limited amount of resources, using a custom
assembler-like language.

Not surprisingly, there are are also attempts to run a Java Virtual
Machine (JVM) on sensor nodes~\cite{javavm_small_devices}.
Squawk~\cite{squawk} is a JVM by Sun Microsystems that runs on Sun
Spots. Obviously such an approach is not suited for low-end sensor
nodes, and also not for time-critical algorithms.

A different approach are macroprogramming frameworks such as
Kairos~\cite{kairos}, Marionette~\cite{marionette}, and
MacroLab~\cite{macrolab}. Instead of writing code for individual
nodes, the whole network is addressed with a single program. This is
generally achieved by providing a script language that is executed
automatically on all nodes, without the need for reprogramming any
node in the network.



%% file: sec_problem.tex
\section{Problem Space}
\label{sec:problem}

\subsection{Heterogeneity}
\label{sec:problem:hetero}

When developing an algorithm library for sensor networks, one must
deal with a great variety of different hardware and software
platforms. \tablename~\ref{table:hardware_platforms} shows an overview
of platforms that were taken into account for the development of the
Wiselib.

\begin{table*}
  \center
  \begin{scriptsize}
    \begin{tabular}{|ll|cccccc|}
      \hline
      Hardware &Firmware/OS&CPU         &Language&Dyn Mem& ROM   & RAM &Bits\\
      \hline
      iSense   &iSense-FW  &Jennic      &C++     & Physical  & 128kB &92kB &32\\
      ScatterWeb MSB&SCW-FW&MSP430      &C       & None      & 48kB  &10kB &16\\
      ScatterWeb ESB&SCW-FW&MSP430      &C       & None      & 60kB  &2kB  &16\\
      Tmote Sky&Contiki    &MSP430      &C       & Physical  & 48kB  &10kB &16\\
      MicaZ    &Contiki    &ATMega128L  &C       & Physical  & 128kB &4kB  &8 \\
      TNOde    &TinyOS     &ATMega128L  &nesC    & Physical  & 128kB &4kB  &8 \\
      iMote2   &TinyOS     &Intel XScale&nesC    & Physical  & 32MB  &32MB &32\\
      GumStix  &Emb. Linux &Intel XScale&C       & Virtual   & 16MB  &64MB &32\\
      \hline
      Desktop PC&Shawn &various     &C++ & Virtual  &unlimited&unlimited&32/64\\
      Desktop PC&TOSSIM&(ATMega128L)&nesC&(Physical)&unlimited&unlimited&(8)\\
      \hline
    \end{tabular}
  \end{scriptsize}
  \caption{Evaluation of potential target platforms. The columns refer to
    the type of microcontroller, the standard operating system, the
    programming language for it, what kind of dynamic
    memory is available, the amount of ROM and RAM, and the bit
    width.}
  \label{table:hardware_platforms}
\end{table*}

The operating systems vary from system-specific implementations such
as iSense and ScatterWeb to generic approaches such as
Contiki, TinyOS, and Linux. The preferred programming languages vary
with the OSs. The iSense firmware has been developed in C++, whereas
the ScatterWeb firmware uses plain~C. TinyOS uses a custom language,
the C extension \textit{nesC}~\cite{nesc03}. Support for dynamic
memory, {\tt malloc()} and {\tt free()}, is only available for some
systems. Using the ScatterWeb firmware, the size of all memory blocks
must be known at compile time, whereas the iSense firmware provides a
full implementation for the C++ operators \texttt{new} and
\texttt{delete}. This is done with the aid of an own memory allocation
implementation. Similar approaches are provided by TinyOS via
\textit{TinyAlloc}, and Contiki via the \textit{managed memory
  allocator} or \textit{memb block memory allocator}. Only the
Linux-based node supports virtual address space for processes. There
are also significant differences in the amount of available memory,
ranging from a few kilobytes to 64 MByte in the GumStix.  Finally, we
must also deal with different bit widths. The Atmel Atmegas are 8-bit
microcontrollers, the MSP430 are 16-bit microcontrollers, whereas the
rest are 32-bit microcontrollers. There are a number of challenges
stemming from the nodes' properties and capabilities. These became
additional library requirements.

\smallskip\noindent\textbf{Limited Memory.}
The algorithms may run on tiny microcontrollers for which the provided
memory is very limited. On the one hand, this affects the ROM. The
generated code for an algorithm must be as small as possible to fit into memory. On the other hand, the RAM
is affected. Routing tables, for example, cannot be arbitrarily long
so as not to exhaust the limited main memory. Additionally, the node
representation that is used for storing the neighborhood must be as
small as possible, but must also meet the demands of the used
algorithms. At the same time, when running on a node with plenty of
memory, performance gains can and should be achieved by employing more
advanced data structures.

\smallskip\noindent\textbf{Physical Dynamic Memory.}  The availability of dynamic memory
  allocation is already a big step forward, allowing for efficient
  data structures. However, most implementations only provide physical
  addresses, and some are even unable to join adjacent freed memory
  blocks. Shifting of pages to join free blocks is impossible on all
  nodes with physical memory. Even a simple vector implementation
  with $O(\log n)$ amortized insertion time would leave behind a trail
  of $O(\log n)$ free blocks of various sizes.  Therefore, data
  structures must be carefully re-analyzed to take these special
  considerations into account.

\smallskip\noindent\textbf{Limited Computation Power.}
Because algorithms may run on small microcontrollers, efficiency plays
an essential role. Examples are message reception in an interrupt or
iterating over a neighbor table to select the next routing node. This
also constrains the Wiselib not to enforce the use of slow operations
(such as excessive pointer indirection) through the provided
framework.

\smallskip\noindent\textbf{Compiler Variance.}
Our library must run on multiple hardware platforms. Different compiler 
versions must be supported, so
it is important that only standard features of the selected
programming language are used.

\smallskip\noindent\textbf{Data Access.}
When accessing data at arbitrary locations in memory, alignment
problems can occur. For example, a cast of a 16bit integer works for
both MSP430 and Jennic, when it starts at an even address. But when it
starts at an odd address, it fails on both platforms. However, a cast
of a 32bit integer works on all even addresses on a MSP430, but for
Jennic only on quad-byte boundaries.

Moreover, when exchanging data in heterogeneous systems, the byte
order must be taken into account, because some systems are big endian,
whereas others are little endian.

\subsection{C++ in Embedded Systems}
\label{ssec:problem:cpp}

The Wiselib must cover all of the previously mentioned hardware and
software platforms; the latter are developed in different programming
languages. Hence, an appropriate programming language must be
found. We chose C++~\cite{stroustrup2000cpp}, because it combines
modern programming techniques with the ability of writing efficient
and performant software. The use of C++ in embedded systems has
already been evaluated~\cite{tr_cpp_performance}. Based on this report
and own evaluations, we selected a subset of the language to be used
in the Wiselib.

C++ allows modern OO designs. Object-Oriented programming is standard
on the desktop for quite some time by now, and has proven to ease the
development of complex systems. Moreover, C++ is a fully typesafe
language. This speeds up the development process, as it catches type
errors at compile time. Given the tediousness of debugging on sensor
nodes, this is a huge achievement.

The most important language feature for the Wiselib are
templates~\cite{vandevoorde03templates,alex01moderncpp}. Templates can
be used to develop very efficient and flexible applications.  The
basic functionality of templates is to allow the use of generic code
that is fully resolved by the compiler when specific types are
given. Thereby, only the code that is actually needed is generated,
and methods and parameters as template parameter can be accessed
directly. We use the well-established technique of template-based ``concepts'' and ``models'', where the former are not specified as actual code, but rather as formal specifications in documentation. It lists the required and provided types, as well as member
function signatures. Models are implementations of concepts, using template specializations, without any inherent runtime overhead. Both concepts and models allow for polymorphism, including multiple inheritance. These techniques
are used successfully in standard C++ 
libraries, such as the STL, Boost~\cite{boost}, and CGAL~\cite{cgal}. The Wiselib employs these methods in the same manner, i.e., using standard compiler features without custom additions.
 

Another basic feature in C++ is virtual inheritance. When declaring a
method as \virtual, the compiler has to generate a vtable consisting
of function pointers to the appropriate methods.  Whenever such a
method is called, it has to be looked up in the \vtable\ first,
thereby requiring pointer indirection. This leads to an increase of
both program memory and run-time, and makes some compiler
optimizations impossible. Hence, we do not use virtual inheritance in
the Wiselib. We substitute this feature by templates.

Two more features that are not used in the Wiselib are run-time type
information (RTTI) and exceptions. Both result in significant runtime
and code-size overhead, as already shown in
\cite{tr_cpp_performance}.





There are C++ compilers available for all of our target platforms. See
\tablename~\ref{table:platforms_cpp} for an overview. Some platforms
lack support for libstd++, which includes the operators
{\tt new} and {\tt delete}. The STL is also not available
everywhere. All compiler support the C++ features we build upon, i.e.,
template and member specializations.
\begin{table*}[t]
  \center
  \begin{scriptsize}
    \begin{tabular}{|l|cccccc|}
    \hline
    Architecture&Compiler  &Binary &Base &libstdc++&Basic C++ Syntax&Templates\\
    \hline
    Jennic      &ba-elf-g++&$\surd$&GCC 4.2.1&$\surd$  &$\surd$     &$\surd$\\
    MSP430      &msp430-g++&-      &GCC 3.2.3&-        &$\surd$     &$\surd$\\
    ATMega128L  &avr-g++   &-      &GCC 4.1.2&-        &$\surd$     &$\surd$\\
    Intel XScale&xscale-g++&$\surd$&GCC 3.3.1&$\surd$  &$\surd$     &$\surd$\\
  \hline
  \end{tabular}
  \end{scriptsize}
  \caption{Availability of C++ compilers for selected platforms.}
  \label{table:platforms_cpp}
\end{table*}


All compilers are based on GCC, and
thus there are no considered drawbacks from compiler
incompatibilities. There are some minor limitations due to the missing
libstdc++ on some systems, which have no impact on the Wiselib.

%% file: sec_wiselib.tex
\section{The Wiselib}
\label{sec:wiselib}

The core design pattern for the Wiselib are generic programming
techniques that are implemented using C++ templates. The basic idea is
to pass the important functionality as template parameters to an
algorithm: implementations of OS specific code, and data
structures. Hence, it is possible to compile an algorithm exactly for
the current needs.

\subsection{Architecture}

\begin{figure}
  \centering
  \includegraphics[width=\columnwidth]{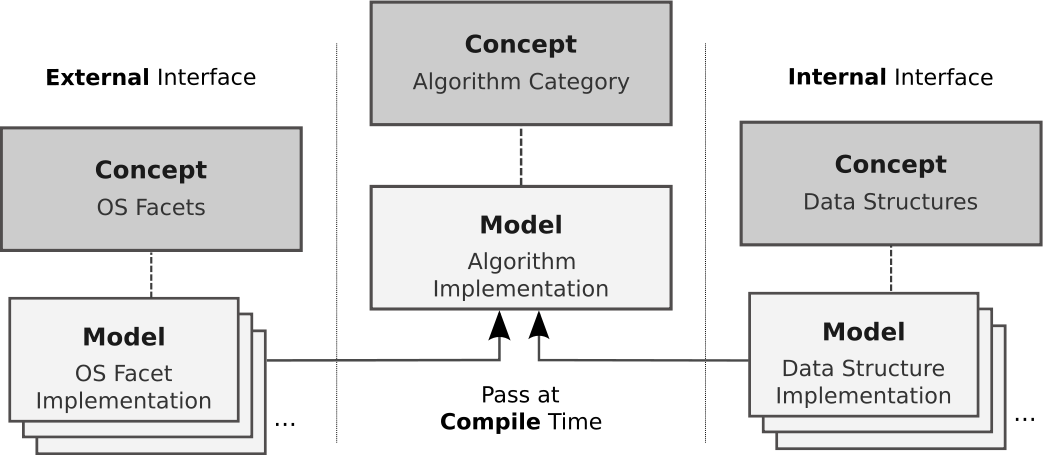}
  \caption{Wiselib Architecture.}
  \label{fig:wiselib_arch}
\end{figure}

The fundamental design principle of the Wiselib consists of
concepts and models, which have already been discussed in
Section \ref{ssec:problem:cpp}. We feature an architecture with three
main pieces: algorithms, OS facets, and data structures. The idea is
shown in \figurename~\ref{fig:wiselib_arch}.

First of all, there are concepts for algorithms. There is one concept
per category, whereby a category groups algorithms by their basic
functionality, e.g. routing or localization. Any algorithm model
implements one or multiple concepts, and is basically a template
expecting various parameters. These parameters can be both OS facets
and data structures.

OS facets represent the connection to the underlying operating system
or firmware---for example, concepts for a radio or timer
interface. Thus, the facets provide a lightweight abstraction layer to
the OS. Note that the facets are merely type definitions and wrapper
functions, they are supposed to contain no replication of OS
functionality.

With the aid of data structures, an algorithm can scale to the
platform it is compiled for. For instance, static data structures can
be passed on tiny platforms without dynamic memory management, whereas
highly dynamic and efficient data structures are passed on powerful
microcontrollers or desktop PCs.


\subsection{External Interface}

The ``external interface'', consisting of OS facets, represents the
connection to the underlying OS. Implementations of these facets are
passed to an algorithm as template arguments. The compiler should
mostly be able to directly resolve such calls to the OS. For example,
when registering a timer can be done using one line of code, it is
implemented as an inline function in the appropriate timer model. Hence, the
result would be a direct call to the OS function, and thus there would
be no overhead, neither in code size nor in execution time. In C-based
operating systems (we see TinyOS in this group), the OS facets have to
provide a translation between C++ member function calls and C function
calls, and they have to convert C++ members to C callback
pointers. This is where an actual price of generality has to be
paid. Fortunately, as we report in Section~\ref{sec:results}, this
price is very low.

Several models of the same concept for an OS facet can also be made
available, each with its own advantages for special purposes. The user
can pass the best available model to an algorithm at
compile time, without extra overhead.


An example for a model of the OS facet ``radio'' is as follows. It is for the
C++-based iSense firmware:
\begin{lstlisting}[language=C++]
template<...> class iSenseRadioModel {
static int send(Os *os, id_t id, size_t len, data_t *data)
{ os->radio().send( id, len, data, 0, 0 ); }
\end{lstlisting}
The example shows the implementation of a simple send method offered
by a radio model. Since it is only one function call, it can be
directly resolved by the compiler without generating any overhead.

\smallskip\noindent\textbf{Concept Inheritance.} The above example of
the radio's send() method with destination address and payload is
defined in the basic radio concept. Routing algorithms, for example,
which do only need to send and receive messages without any further
information such as RSSI values, or requirements such as reliable
delivery can use implementations of this concept.

We also allow for concept inheritance, so that the basic radio
concept can easily be extended. If an algorithm needs access to
RSSI (or LQI) values, a derived concept can be used. It 
extends the basic one with a receive method that provides additional
values. 

\smallskip\noindent\textbf{Stackability.} A major design aspect for the
radio concept is stackability, i.e., the possibility to build a
layered structure of multiple radios. The topmost layer is not
aware to which and how many layers it is connected. The big advantage
of this approach is that we can build a
``virtual radio'' that runs on top of a radio model, and is passed to
an algorithm in its radio template parameter. Doing so, we can easily
implement an algorithm for heterogeneous sensor networks. It is even
possible to communicate between nodes that use different kinds of node
IDs---because the virtual radio hides the real node
addresses and provides, e.g., generic 128~bit addresses.

Another possibility is to hide a complete routing algorithm behind an
OS facet. For example, when writing out debug messages, this happens
generally to the UART. But by passing another model, we can
forward debug messages over a routing algorithm to a gateway,
where all these messages are collected. The topmost algorithm does not
need to be aware of the model it works on---it must only use the
appropriate concept.

\smallskip\noindent\textbf{Message Delivery in Heterogeneous Systems.}
Another problem that is addressed using our software design is message delivery
in heterogeneous networks. There are basically two problems that occur:
different byte-order, and differences in alignment handling. Byte order issues
are solved by sticking to network byte order in messages. Alignment is
addressed via template specialization. We provide a serialization class that
provides generic \texttt{read} and \texttt{write} methods for all data types.

\subsection{pSTL} Not all of our target systems provide dynamic memory
allocation. To our knowledge, no variant of the STL fulfills our
requirements: not using libstdc++, new/delete, exceptions, and
RTTI. 

Consequently, we provide the pSTL, an implementation of parts of the STL that
does neither use dynamic memory allocation nor exceptions nor RTTI. We ensure
that each of the provided data structures works on each supported hardware
platform. At the moment, implementations for {\tt map}, {\tt vector}, and {\tt
list} are available. Naturally, the pSTL will grow with increasing demand.

\subsection{pMP}
For many tasks in embedded systems, multi-precision arithmetic is needed,
e.g. for cryptographic and data aggregation purposes. Currently there exist a number
of software libraries that implement big-number operations, e.g., gnuMP~\cite{gmp}. Such
libraries heavily rely on dynamic memory allocation to represent big-numbers and carry out the operations.
Moreover, to achieve performance speedups, highly optimized assembly code is used, taking advantage of
specific hardware instructions.
Unfortunately, the hardware types used in WSN platforms (e.g., ATMEGA, Jennic) support
neither dynamic memory allocation nor the specific hardware instructions used by gnuMP and other libraries.
Hence it is very difficult to port such implementations to our platforms, if not impossible at all.

Therefore, we provide the pMP, an C-based implementation of big-number operations that
does not use dynamic memory allocation. Of course such
a library cannot be compared in terms of efficiency with gnuMP, but it is the
only one available currently. In particular, it implements some basic
operations like xor, shiftleft and modulo multiplication operations which are
required for elliptic curve cryptography. It is certain that the pMP will grow
regarding future needs.

\subsection{Algorithm Support}

The central piece of the Wiselib are the algorithms. They are
grouped into categories, see Section~\ref{sec:intro}. Algorithm implementation can belong to several categories,
which is common for cross-layer algorithms.

Each algorithm class consists of a concept for the
algorithm itself, and some concepts for the data structures
that are typically necessary for this class. This decouples the
algorithm logic, which is invariant over different platforms, from
data storage, which heavily changes when an algorithm is ported to a
platform of different characteristics.

The benefit of having a well-defined algorithm interface is that
algorithms are easily interchanged for testing purposes, ideally this
is done by simply altering a class name in the initialization
code. The second---much more important---benefit is that an algorithm
developer can start coding by copy-and-paste, instead of having to go
through a design phase. Such a design phase can be quite lengthy, if
the goal is to achieve maximal portability. Until now, theoreticians wishing to
evaluate high-level algorithms often found it hard to develop for
embedded devices: this lowers the bar considerably.

Providing a diverse set of data structure implementations serves the
goal of scalability: For each data structure, e.g., routing tables,
neighborhood cluster maps, and position maps, a set of implementations
matching the span of platforms is provided. For low-end architectures
such as the MSP430, structures are needed that use static storage
whose size is known at compile-time. Such structures will inevitably
be inefficient in terms of runtime. For high-end architectures using
Xscale processors or simulation environments, highly optimized data
structures with dynamic memory management and huge memory overhead can
be employed, resulting in high efficiency. It is even feasible to
utilize the STL. The choice of data structures has no impact on the
algorithm code, and can simply be configured at algorithm
initialization. This results in algorithms that not only scale down to
very limited devices, but also scale up to powerful nodes, utilizing
all the available resources on them.


%% file: sec_case_study.tex
\section{Case Study: Secure Routing Algorithms}
\label{sec:case_study}

We show the benefits of C++ and template-based design by presenting
two examples: routing and cryptography algorithms. First we present
either of the approaches as a single concept. Then we show how
easily individual implementations can be combined to generate secure
routing algorithms.

\smallskip\noindent\textbf{Routing Algorithms.}
When designing a concept for an algorithm class, one wishes to cover
all kinds of special case, while staying as generic as possible. This
is because each method in the concept must be implemented by each
model. Hence, our concept for a routing algorithm consists of only six
methods.

First, we need a method for setting the pointer to the OsModel that is
needed when calling static member functions from the External
Interface. Then we have two methods for enabling and disabling the
routing algorithm, which is useful when the routing should only be run
in certain points in time, for example for energy-saving issues. Next,
a potential user of the routing algorithm must be able to register and
unregister a callback for message reception. At last, there is the
method for sending messages to other nodes in the network. The Routing
Concepts specializes the Radio Concept, so that routing algorithms can
be used as virtual radio interfaces for other algorithms. The concept
looks as follows:
\begin{lstlisting}[language=C++]
concept Routing {
  void set_os(OsModel* os);
  void enable(void);
  void disable(void);
  void send(node_id_t receiver, size_t len, data_t* data);
  template <class Callee, void (Callee::*Method)
                       (node_id_t, size_t, data_t*)>
  int reg_recv_callback(T *obj_pnt);
  void unreg_recv_callback(int);
};
\end{lstlisting}

\smallskip\noindent\textbf{Cryptography.}
Adapting cryptographic algorithms to embedded systems is a 
difficult task due to resource limitations. Unlike the routing case,
we avoid covering all special cases of crypto algorithms. We
provide a simple concept with algorithm implementations that
will be viable solutions for the tiny sensors.

Our generic concept for a crypto algorithm consists of five methods.
We provide methods for key setup, encryption and decryption of data
blocks. The concept looks as follows:

\begin{lstlisting}[language=C++]
concept Crypto {
   void set_os(OsModel* os);
   void enable(void);
   void disable(void);
   void key_setup(node_id_t, data_t* key);
   void encrypt(data_t* in, data_t* out, size_t length);
   void decrypt(data_t* in, data_t* out, size_t length);
};
\end{lstlisting}

\smallskip\noindent\textbf{Secure Routing.}
In this section, we describe how the individual routing and cryptographic
implementations can be combined to result in secure routing
algorithms. Note that any available routing implementation can be
combined with any available crypto algorithm without a single change
in their code.

We therefore implement the routing concept, and accept a routing
algorithm and a crypto algorithm as template parameters. Internally,
we only use the passed types. For example, when the secure routing is enabled,
it in turn enables the routing and crypto algorithm. When a message is
sent, it first encrypts the passed bytes, and then passes the
encrypted data to the routing algorithm. Then, when a message is
received at the destination, it is first decrypted, and then passed to
the registered receivers. The secure routing looks then as follows:

\begin{lstlisting}[language=C++]
template<typename Routing,
         typename Crypto>
class SecureRouting {
   void set_os(OsModel* os);
   [...] // all methods described in the routing concept
   void unreg_recv_callback(int);
   Routing routing_;
   Crypto crypto_;
};
\end{lstlisting}


Since it implements the routing concept, it can be passed and used by
any application that deal with routing algorithms. However, the
process of both encryption and decryption is completely transparent.


%% file: sec_results.tex
\section{Experimental Results}
\label{sec:results}

In order to demonstrate the efficiency of our generic approach,
we ran different experiments on supported platforms. 
We evaluated two main
parts of the Wiselib: First, the overhead of the connection to the
underlying OS; second, properties of implementations of a first set of
algorithms.

\subsection{External Interface}

We tested the performance of Wiselib system calls compared to
native OS calls on three different platforms. The results are shown in
\tablename~\ref{table:external_iface_performance}.

\begin{table*}[b]
  \center
  \begin{scriptsize}
    \begin{tabular}{|l|c|c|c|c|c|c|c|c|c|}
      \hline
      &\multicolumn{3}{|c|}{iSense}&\multicolumn{3}{|c|}{Contiki}&
                                             \multicolumn{3}{|c|}{ScatterWeb}\\
      & Native   & Wiselib  &Cost & Native  &Wiselib  & Cost   & Native & Wiselib & Cost\\
      \hline
Read ID     &  2$\mu s$&  2$\mu s$&0\%&$<$1$\mu s$&$<$1$\mu s$&0\%&$<$1$\mu s$&$<$1$\mu s$&0\%\\ 
Send Message&282$\mu s$&282$\mu s$&0\%&336$\mu s$ &345$\mu s$ &3\%&898$\mu s$&921$\mu s$&3\%\\ 
Set Timer   &135$\mu s$&141$\mu s$&4\%&  77$\mu s$&100$\mu s$ &30\%& 20$\mu s$&43$\mu s$&115\%\\ 
      \hline
    \end{tabular}
    \caption{Performance costs of Wiselib calls compared to native OS
      calls.}
    \label{table:external_iface_performance}
  \end{scriptsize}
\end{table*}

OS calls that are short enough to be directly inlined by the compiler,
such as sending a message on iSense platforms or reading the node ID
in Contiki do not have any overhead. However, other parts in the OS
connection produce a small overhead due to an additional layer of
indirection. This is mainly because of incompatibilities between C
function pointers and C++ member function pointers, and a required
translation between them. But as shown in the performance evaluation,
this overhead is very small---if at all, then only in terms of
microseconds. Similar delays would also be produced by alternative
approaches, but by using C++ and templates the compiler is able to
remove this overhead wherever reasonable. This is possible due to the
implicit inline declaration of methods.

Time efficiency is only one performance measure; the other is
code space. We evaluated the needed size for the two OS facets radio
and timer for different platforms. The results are shown in
\tablename~\ref{table:external_iface_byte_overhead}.

\begin{table*}[t]
  \def\db#1#2{\makebox[2.3em][r]{#1}+\makebox[1.5em][r]{#2}}
  \center
  \begin{scriptsize}
    \begin{tabular}{|l|c|c|c|}
      \hline
            &iSense       &Contiki      &ScatterWeb\\
      \hline
      Radio &\db{856}{240}&\db{428}{ 72}&\db{316}{40}\\
      Timer &\db{868}{240}&\db{352}{210}&\db{270}{80}\\
      \hline
    \end{tabular}
    \caption{Code-size overhead of OS facets. Shown is ROM (.text) and
      RAM (.bss + .data) in bytes.}
    \label{table:external_iface_byte_overhead}
  \end{scriptsize}
\end{table*}

Because the concepts for radio and timer were kept simple, each
implementation required at most a few hundred lines of code. This led
not only to a slight structure, but also enhanced maintenance
issues. In addition, even the integration of a completely new platform
can be done without too much effort.

Especially the facets for the ScatterWeb platform show a small amount
of overhead of less than 600 bytes in ROM, and 120 bytes in RAM. 
Even the 1.7kB of iSense are tolerable, since it is a 32bit-platform
with corresponding overhead in machine language instructions. 

An important factor when estimating the code-size overhead is that it
is constant, and thus do not grow with the integration of further
algorithms. The interfaces also provide a powerful abstraction of the
underlying OS, facilitating implementations of many additional algorithm categories.

\subsection{Algorithms}

We implemented different algorithms for the routing concept: DSDV,
DSR, a simple tree routing, and a flooding algorithm. Each algorithm
has been compiled for, and tested on each supported
platform. \tablename~\ref{table:casestudy:codesize} shows the
resulting code sizes and initial RAM usage for the several platforms.

\begin{table*}
  \def\db#1#2{\makebox[2.3em][r]{#1}+\makebox[1.5em][r]{#2}}
  \center
  \begin{scriptsize}
   \begin{tabular}{ | l | c | c | c | c | c | }
     \hline
  & \multicolumn{2}{|c|}{16-bit OS} & 32-bit OS & \multicolumn{2}{|c|}{Simulators} \\
 Algorithm  & Contiki      & ScatterWeb   & iSense       & Shawn      & TOSSIM \\
  \hline
  DSDV       &\db{1446}{72} &\db{1466}{72} &\db{4776}{136}&\db{4351}{4}&\db{19146}{4}\\
  DSR        &\db{1964}{338}&\db{1716}{238}&\db{5396}{356}&\db{6918}{4}&\db{20845}{4}\\
  Tree       &\db{ 920}{16} &\db{ 724}{14} &\db{4060}{ 24}&\db{2974}{4}&\db{9946}{4}\\
  Flooding   &\db{1122}{50} &\db{ 762}{34} &\db{2864}{ 68}&\db{2260}{4}&\db{10192}{4}\\
  \hline  
    \end{tabular}
    \caption{Evaluation of code size as 
      ROM size (.text) and RAM size (.bss + .data) in bytes.}
    \label{table:casestudy:codesize}
  \end{scriptsize}
\end{table*}

It is clearly visible that our algorithm implementation perfectly fits
into the target platforms, as the impact of the generality of the code
is very low, in terms of both code and memory. However, the given code
sizes show only the pure demand of the algorithm---without considering
the external interface.

Each of the routing models can also be combined with a crypto
algorithm---as shown in Section~\ref{sec:case_study}. The first point
of interest is the overhead of multiple layers of algorithms are.
We estimated the average latency by the Wiselib
layers. The experiments were held on the iSense platform. The latency
was measured as the average of 200 message exchanges: a) through a
dummy routing algorithm and a dummy routing algorithm combined with a
dummy crypto algorithm and b) through a DSDV routing algorithm and a DSDV
routing algorithm combined with a dummy crypto algorithm. We conclude
that stack latency overhead is minimal, as shown in \tablename~\ref{table:stack_latency}.

\begin{table*}[t]
  \center
  \begin{scriptsize}
    \begin{tabular}{|l|c|c|c|c|}
      \hline
      &Dummy Routing &Dummy Routing,&DSDV Routing & DSDV Routing,\\
      &              &Dummy Crypto  &             & Dummy Crypto\\
      \hline
      Latency & 6.08 msec  & 6.09 msec &6.72 msec &6.75 msec\\
      \hline
    \end{tabular}
    \caption{Stack latency in Wiselib (measured on the iSense
      devices).}
    \label{table:stack_latency}
  \end{scriptsize}
\end{table*}

As a second experiment regarding the combination of routing and crypto
algorithms, we estimated the run-time of a crypto algorithm (Elliptic
Curve Integrated Encryption Scheme) through Wiselib for various
platforms, and we compared it with that of TinyECC\cite{tinyecc} in \tablename~\ref{table:crypto_on_different_platforms}. We did not focus on
optimizing the code; that is why TinyECC runtime is generally
faster. However, our algorithm can be executed on a variety of
platforms.

\begin{table*}[b]
  \def\db#1#2{\makebox[2.3em][r]{#1}+\makebox[2.3em][r]{#1}}
  \center
  \begin{scriptsize}
   \begin{tabular}{ | l | c | c | c | c | c | c | }
     \hline
  & \multicolumn{2}{|c|}{TinyECC optimized} & \multicolumn{2}{|c|}{TinyECC } & \multicolumn{2}{|c|}{Wiselib} \\
 Hardware  & Encrypt      & Decrypt & Encrypt      & Decrypt & Encrypt      & Decrypt \\
  \hline
  TelosB       &6.53sec &4.25sec &84.9sec&42.73sec&114.78sec&56.02sec\\
  MicaZ        &3.9sec & 2.6sec & 61.4sec & 31.87sec  & 118.4sec & 57.84sec\\
  Tmote Sky    & 3.27sec & 2.12 sec & 42.55sec & 21.41sec  & 115.98sec & 56.91sec  \\
  iSense       & - & - & -   & -       & 22.9sec & 11.84sec\\
  ScatterWeb   & - & - & -  & -        & 102.93sec & 50.42sec\\
  \hline  
    \end{tabular}
     \caption{Comparison between Wiselib and TinyECC, for
        encryption/decryption runtime.}
     \label{table:crypto_on_different_platforms}
  \end{scriptsize}
\end{table*}

Also, with the aid of template specializations---as also used in
message delivery---code can be optimized and adapted for certain
platforms. Depending on the compilation process, the compiler can
select exactly the code that fits best for the current platform. For
example, when an algorithm is compiled for iSense, the AES hardware
could be used for the crypto routines.


%% file: sec_access.tex
\section{Accessing the Wiselib}
\label{sec:access}

There are different demands for the users of the
Wiselib. \textit{Application developers} are interested in
stable algorithms that were thoroughly tested for all supported
platforms. They do not contribute own implementations to
the Wiselib; instead, they only integrate existing algorithms in their
applications. \textit{Algorithm developers} on the other hand
contribute code to the Wiselib. Algorithms may be under development
and can not be ensured to run on each platform.

We therefore provide two distributions: \textit{Stable} and
\textit{Testing}. The former contains only algorithms that were run
through different tests, particularly for each supported
platform. Concepts that are implemented for the stable distribution
are also expected not to be changed anymore, if not strongly
needed. In contrast, the testing distribution contains newly
implemented algorithms. They may not be tested on each platform---in
particular since not each algorithm developer has each platform
available. This can also lead to changes in concepts, when it is
noticed that not all platforms can be covered satisfactorily. In
general, the objective here is to release early, and release often.

The Wiselib can be accessed under
\url{http://wisebed.eu/wiselib}. There is a Wiki available that
contains documentation. In addition, there is also a Trac running to
report software bugs and collect suggestions for improvement.


%% file: sec_conclusion.tex
\section{Conclusion and Future Work}
\label{sec:conclusion}

In this paper, we have introduced our generic algorithm library for
wireless sensor nodes, the Wiselib. It is aimed at allowing algorithm
researchers to quickly implement distributed algorithms on actual
sensor nodes. The implementation process requires no deep
understanding of the target platform, as the library provides a
unified API that abstracts the technical details. Unlike all other approaches with the same goal, or at least
the ones we are aware of, Wiselib algorithms suffer next to no runtime or
memory overhead from the generality. 

The Wiselib is written in standard ISO C++, using advanced OO
techniques to encapsulate the operating system and to allow complex OO
architectures that can be fully resolved by an optimizing
compiler. Specifically, the Wiselib makes heavy use of templates, as
they are resolved at compile time, leaving no binding efforts to
runtime. Certainly, generality does not allow to provide highly optimized code. 
Fortunately, our open design allows to provide such hardware specific 
optimizations without hindering the generality of the algorithm implementation.
This is extremely important since algorithm development can be decoupled
from application development where platform specific optimizations are performed.

We demonstrate the effectiveness of the Wiselib by implementing a
number of routing algorithms and cryptography algorithms. 
We show that the produced code is very lean and it works on a large variety of sensor platforms. 
The library allows us to easily
stack different types algorithms with almost zero overhead. We build upon
this feature and demonstrate the ability 
to interchange algorithms without affecting the operation of other
algorithms at different stack level. These features essentially provide endless
possibilities to application developers as more algorithms and algorithmic concepts
are introduced in Wiselib.

We expect the Wiselib to grow much beyond the current state, and to
become a standard tool for WSNs in the near future. We also wish to
look into other categories of algorithms such as MAC layer protocols,
energy saving schemes and topology control protocols.

